\documentclass[11pt,twoside]{article}
\usepackage{asp2010}
\resetcounters

\begin{document}

\title{The Cluster Birthline and the formation of stellar clusters in M33}
\author{Edvige~Corbelli$^1$, Carlo~Giovanardi$^2$, and Marco~Grossi$^3$
\affil{$^1$INAF-Osservatorio Astrofisico di Arcetri, L. E. Fermi, 5, 50125-Firenze, Italy}
\affil{$^2$INAF-Osservatorio Astrofisico di Arcetri, L. E. Fermi, 5, 50125-Firenze, Italy}
\affil{$^3$CAAUL, Observatorio Astronomico de Lisboa, Tapada de Ajuda, 1349-018 Lisboa, Portugal}}

\begin{abstract}

We present a new method to analyze the IMF at its high mass end in young stellar 
clusters,
which rely on two integrated observables: the cluster bolometric and H$\alpha$ luminosity.
Using several cluster samples selected in M33 we show that a stochastically sampled universal 
IMF is in better agreement with the data than a truncated IMF whose maximum stellar mass
depends on cluster mass. We also discuss the possibility that a delayed formation of massive stars
is taking place in low density star forming regions as an alternative to a strong leakage of
ionizing photons from HII regions of young luminous clusters.

\end{abstract}

\section{Introduction}

Extragalactic surveys of young stellar clusters in the Local Universe complement Milky Way studies 
because star formation can be traced in different environments: from spiral arms to interarms,
from galaxy centers to galaxy outskirts, from metal poor to metal rich disks, from dwarfs to 
more massive spirals. Even though the physical scales which can be spatially resolved will never  
be as small as in galactic  
surveys, extragalactic studies benefit from the larger database available.
Having a statistically meaningful ensemble of clusters is crucial when testing 
the Initial Mass Function (hereafter IMF), given the fact that the IMF is a  probabilistic 
function describing the 
distribution in mass of stars at their birth. This is especially true when studying
the upper end of the IMF, since massive stars are rare and short 
lived events. The Milky Way disk is not the ideal laboratory for investigating the birth rate of 
massive stars due to their paucity and to the the high dust extinction along the line of sight
which prevent searches over large areas. 
Signatures of massive stars are used to infer the rate of star formation in galaxies, and hence, 
even though most of the stellar mass is in small mass stars, it is very important to investigate
the upper end of the IMF to correctly draw any conclusion on the star formation rate. 
M33, a Local Group galaxy, is an ideal candidate for studying the
formation of young stellar clusters. Its distance is only 840~kpc, its disk is undisturbed,
it has a low inclination (55~\deg) respect to our line of sight, 
a low dust content, and a higher star formation rate per
unit area compared to other Local Group members such as M31. 
Using Spitzer data, \citet{2007A&A...476.1161V}
have cataloged infrared sources  in M33  which span over 4 order of magnitude in luminosity
and recovered sources as faint as embedded B2-type stars. We shall use  the 24~$\mu$m Spitzer
maps of M33 to select stellar clusters which have an infrared counterpart due to
dust in the surrounding interstellar gas, and hence likely to be young cluster candidates. 
 
\section{The cluster birthline}

\begin{figure}[ht!]
\includegraphics[bb=18 104 480 550,scale=0.7]{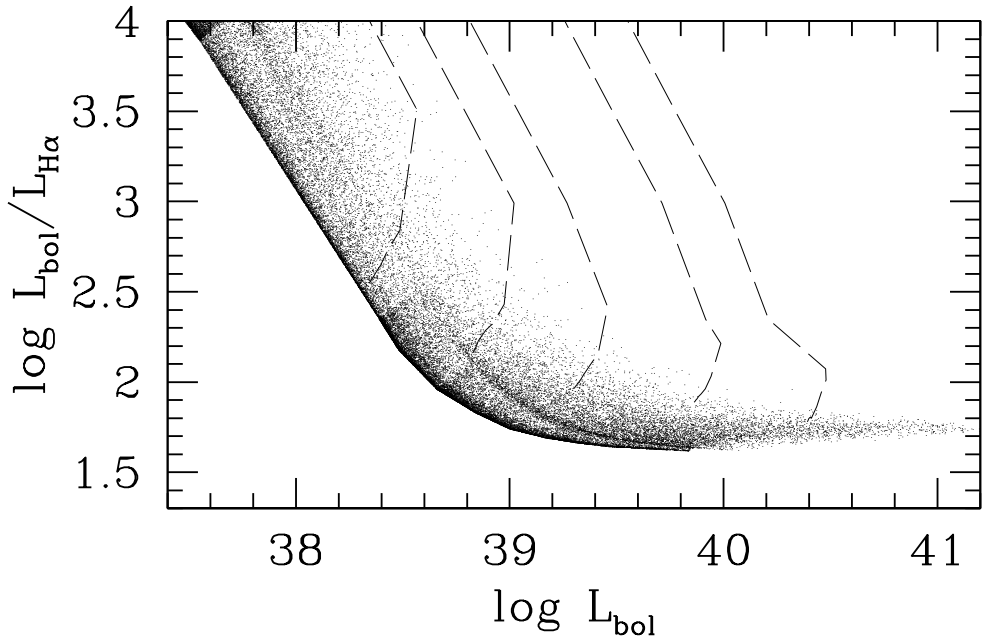}
\caption{The distribution of the ratio of the bolometric-to-H$\alpha$
luminosity versus the bolometric luminosity for newly born clusters simulated
for the stochastically sampled universal IMF extending up to 120~M$_\odot$. 
Vertical dashed lines
mark evolutionary sequences, i.e., aging moves clusters upward from the simulated
birthline because H$\alpha$ luminosities fade away more rapidly than
bolometric luminosities. 
\label{fig:1}
}
\end{figure}

Rather than resolved star counts we shall investigate integrated properties of star 
clusters which can be used to infer their massive stellar population. Low mass stars 
determine the stellar cluster mass because the slope
of the IMF is steeper than $-2$. The cluster bolometric luminosity and the ionizing photon
rate, as measured by the extinction corrected H$\alpha$ emission, are instead linked to 
the massive stellar population. For this reason we shall focus on the last two observables
to describe clusters. For individual massive stars, the bolometric and
H$\alpha$ luminosity scale in a rather similar way with stellar mass while for stars less
massive than 20~M$_\odot$ the H$\alpha$  luminosity drops with stellar mass much more
quickly than the bolometric luminosity does. This implies that when the newly born cluster 
is small and very massive stars are lacking, we cannot easily infer the cluster mass or
the star formation rate in the surrounding region from the observed H$\alpha$ luminosity. 
If the
stellar cluster is massive enough that the IMF is fully populated up to its high mass
end, the ratio between the bolometric and H$\alpha$ luminosity is constant and
it does not depend on the cluster mass. Both the bolometric and the H$\alpha$ luminosity
scale linearly with cluster mass. To pin down the effects of the IMF incompleteness, 
which set in as the cluster mass gets below a critical value and affect the cluster 
luminosity, we introduce the concept of cluster birthline \citep{2009A&A...495..479C}. 
We shall call birthline the 
region in the log L$_{bol}$ -- log L$_{bol}$/L$_{H\alpha}$ plane where newborn young 
stellar cluster lie. We show in Figure 1 the cluster birthline for a stochastically 
sampled IMF whose slope at its high mass end is the Salpeter one i.e. 2.35. 
The shape of the cluster birthline does not depend much on the slope 
of the IMF and on the cluster mass function. Its shape is linked to the 
dependence of the bolometric luminosity and ionizing flux of individual stars on
stellar mass as well as to the mass of the most luminous star that can ever be
formed.

The cluster birthline changes if one assumes a cluster-mass dependent
IMF instead of a stochastically sampled universal function. 
\citet[][and references therein]{2006MNRAS.365.1333W} proposed that the maximum possible
mass of a star in a cluster decreases systematically with decreasing cluster mass. This
implies that no high mass stars can ever be formed in intermediate mass clusters.  
For a randomly sampled universal IMF they can occasionally be formed. As long as there 
is enough gas, intermediate mass clusters can make bright ``outliers''. In Figure~1
we show also the cluster birthline for the maximum mass case according to the IMF model
of \citet[][]{2006MNRAS.365.1333W}. The cluster birthline in the maximum mass case has 
higher values of L$_{bol}$/L$_{H\alpha}$ ratio over the range of L$_{bol}$ where O-
and B-type stars start lacking.

An important property of the cluster birthline is that aging moves the clusters above it.
This is because the death of massive stars makes the cluster H$\alpha$ luminosity fade 
away more rapidly than the bolometric luminosity does. For a given IMF the cluster birthline
is a theoretical lower boundary for the  L$_{bol}$/L$_{H\alpha}$ ratio. In fact possible
leakage or dust absorption of ionizing photons in the HII region increases the
value of the observed L$_{bol}$/L$_{H\alpha}$ ratio. Hence if clusters with a well 
determined value of L$_{bol}$ and L$_{H\alpha}$ lie below a theoretical 
birthline the assumed model for the upper end of the IMF needs to be revised.

A stochastically sampled universal IMF implies that the cluster mass is highly 
undetermined when the observed bolometric luminosity is below $10^{40}$~erg~s$^{-1}$.
Results of a numerical simulation shown in Figure~2 emphasize this by showing
the wide range of cluster masses corresponding to L$_{bol}$ values below
$10^{40}$~erg~s$^{-1}$. Stochasticity allows for example a 
$5\times 10^{39}$~erg~s$^{-1}$ luminous cluster to be made by a single 100~M$_\odot$ 
star or by a more massive ensemble of small mass stars distributed according to a fully 
populated IMF up to about 30~M$_\odot$.
 
\begin{figure}
\plottwo{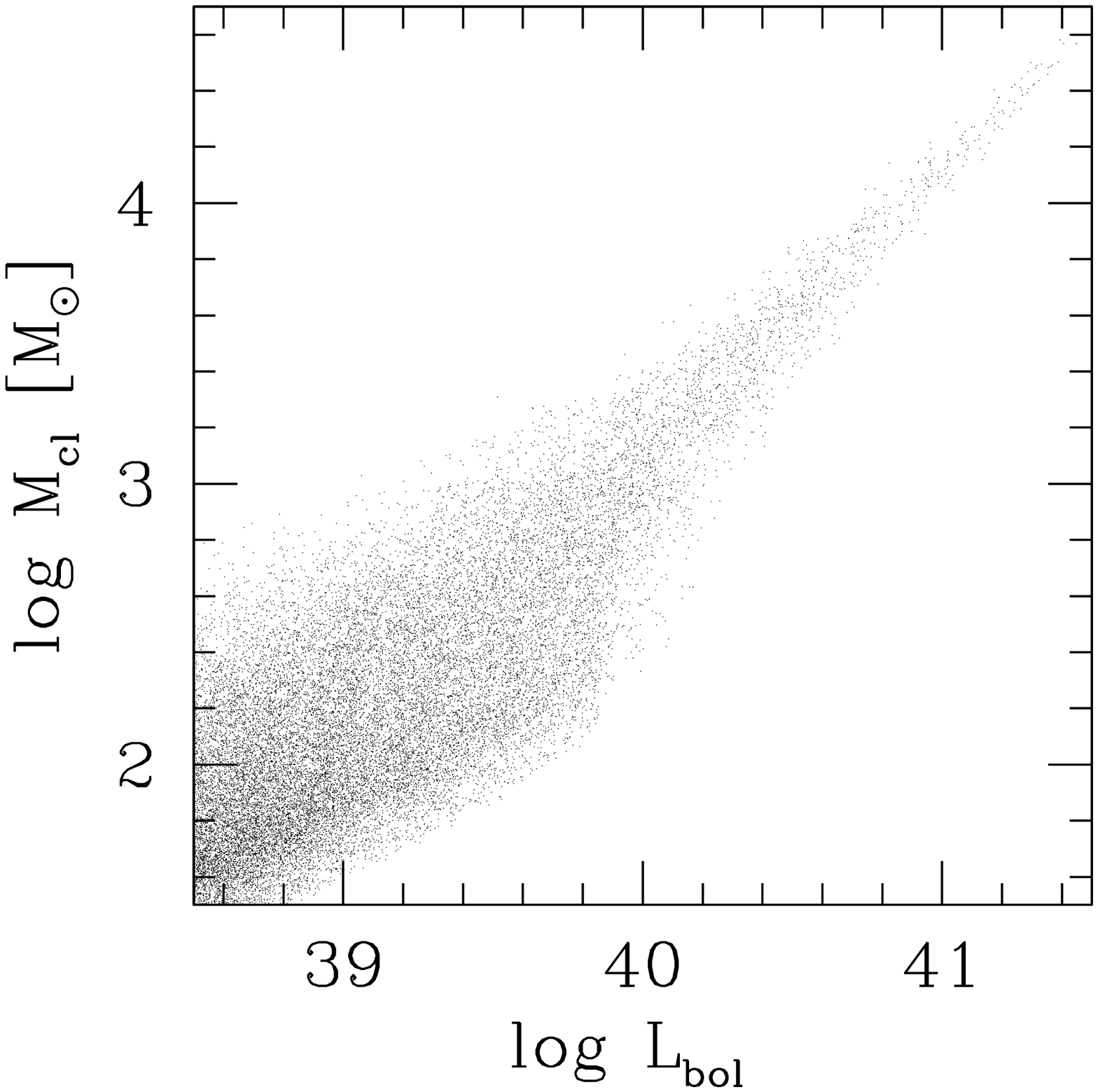}{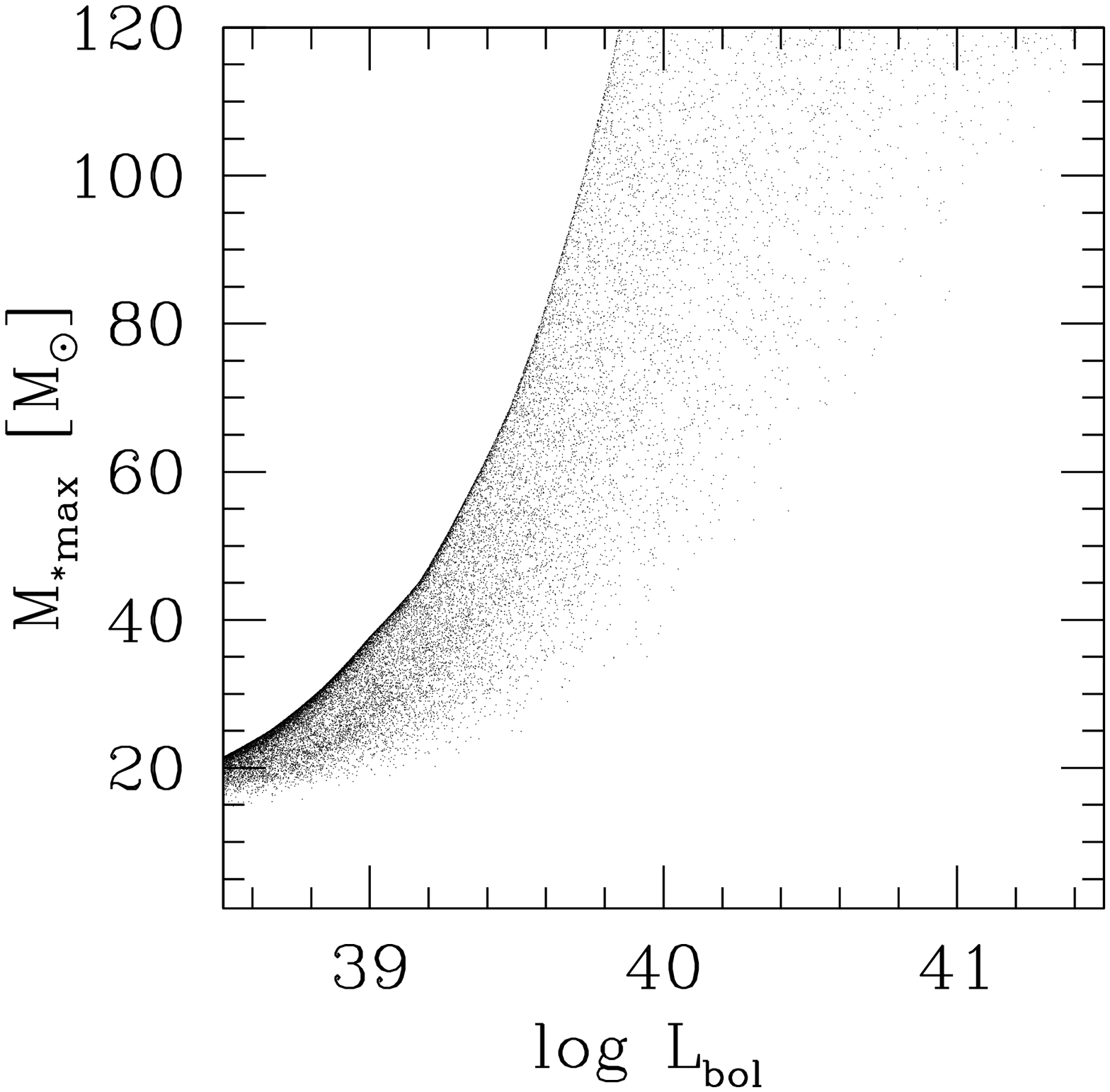}
\caption{Possible values of cluster masses and of the most massive star that forms as
a function of the cluster bolometric luminosity.
The cluster mass here is computed only counting stars more massive than 1~M$_\odot$.
The presence of stars of lower mass will shift log~M$_{cl}$  upward by a constant 
value which will depend on the IMF lower mass cut-off.  
\label{fig:2}
}
\end{figure}

\section{Young stellar clusters in M33}

\begin{figure}
\includegraphics[bb=18 104 480 550,scale=0.6]{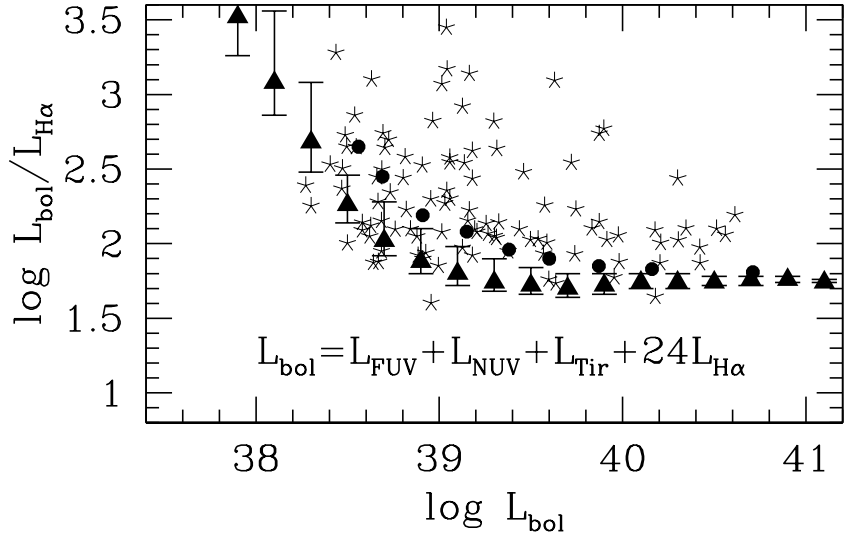}
\caption{Bolometric to H$\alpha$
luminosity ratio versus bolometric luminosity for the infrared selected sample
in M33. Bolometric luminosities have been estimated by adding the emission at
various wavelengths and individual extinction corrections
have been applied to H$\alpha$ luminosities. 
Median values of the theoretical {\it cluster birthline}
for the {\it stochastically sampled universal IMF} (filled triangles) and for the
{\it maximum mass case} (filled dots) are also shown.
\label{fig:3}
}
\end{figure}

In Figure~3 we compare  
the observed values of L$_{bol}$/L$_{H\alpha}$ ratio for an infrared selected sample 
of stellar clusters in M33 with the  cluster birthline for the maximum mass case and  
for the case of a universal  
stochastically sampled IMF. For this last model we plot the expected median values 
of L$_{bol}$/L$_{H\alpha}$ with their relative dispersion. The bolometric
luminosity of clusters in the sample has been computed by adding the far and near UV
luminosities to the total infrared luminosity as inferred from the 24~$\mu$m emission 
\citep{2009A&A...493..453V}. A term which depends on
the H$\alpha$ luminosity has been added to take into account the contribution of
the ionizing part of the spectrum to the bolometric luminosity. We can see that all of
the M33 clusters are compatible with the  stochastically sampled universal IMF. There
is a group of clusters which falls below the birthline relative to the maximum-mass IMF.
This finding suggests that a stochastically sampled universal IMF is more appropriate
than the maximum-mass case. The bolometric luminosity formula that we have
used is valid for young clusters whose unabsorbed light falls mostly in the UV 
bands, which is the case if stars more massive than 10~M$_\odot$ are present.
In the next subsection we will determine the age and bolometric luminosity
of stellar clusters using synthesis model fits to the spectral energy distribution
(hereafter SED).

\subsection{Cluster ages from SED fits}

Multiwavelength observations of 32 young star clusters in M33, from UV to the near 
infrared, has allowed \citet{2010arXiv1006.1281G} to derive stellar cluster properties 
by fitting synthesis models to the cluster spectral energy distribution. Ages, bolometric 
luminosities, masses and extinction have been determined from the best fitting single 
age stellar population burst model with a fully populated IMF.
The selected sample is made of isolated compact sources visible both in infrared, 
optical, and UV images and with a well determined nebular metal abundance.
Cluster ages range between 2 and 15 Myr and bolometric luminosities between 10$^{39}$ and  
2$\times 10^{41}$~erg~s$^{-1}$. Relatively low values for the extinction have been found, 
$0.2<A_v<1.1$, in agreement with what suggested by the Balmer line decrement.
Since we also measured the H$\alpha$ emission from the
HII region around the cluster, in Figure 4 we compare the data for this sample with the 
cluster birthline. We expect younger clusters to lie close to the birthline than
older cluster and instead we seem to find the opposite trend. 
Notice that if 
L$_{bol}$ from SED fits were correct for the whole sample, and hence only a couple of clusters 
have an incomplete IMF, most of the young clusters would be highly underluminous
in H$\alpha$. This means that a high fraction of ionizing photons, sometime very close to 
100$\%$ of them must have leaked out of the HII region. This is in contrast with
previous analysis of the leakage in HII regions of M33 \citep{2000ApJ...541..597H}
and with theoretical models. 

As discussed in
the previous Section, when the cluster bolometric luminosity is below 
10$^{40}$~erg~sec$^{-1}$ incompleteness of the IMF at the high mass end sets in
and should be taken into account. We now compute the bolometric
luminosity of the clusters by adding the emission in the various bands, as in Figure~3,
and only when this is close to 10$^{40}$~erg~sec$^{-1}$ or above it we run SED fits 
to re-determine L$_{bol}$ and the other cluster parameters. Figure~4$(b)$ shows
the results. 
The luminosities predicted by SED model fits using a fully sampled IMF are higher than
those derived by the arithmetic sum for low luminosity clusters,  
as discussed also in \citet{2010arXiv1006.1281G}. This discrepancy warns against an
indiscriminate use of SED synthesis models if stochastic effects are not taken into account.
Detailed SED models 
which take into account stochasticity will help in the future to confirm this finding.
Note that several clusters have now bolometric luminosities below 
10$^{40}$~erg~sec$^{-1}$ and lie along the birthline: the problem of a strong ionizing 
photon leakage is alleviated. Bright clusters however are still underluminous in 
H$\alpha$, especially the young ones. 
For the data shown in Figure~4$(b)$ a standard universal IMF which extends up to 
120~M$_\odot$ implies that on average
about half of the ionizing photons are leaking out of bright HII regions in our sample.
It is worth to mention that despite the
relatively low extinction predicted by the SED models, we expect more light re-emitted in
the IR around these clusters than estimated from M33 Spitzer images using simple dust 
models. This 
suggests that not only ionizing photons but also continuum UV photons are lacking 
in these HII regions, and that dust cannot be responsible for the missing ionizing
photons. 
Notice that M33 is a galaxy with a high diffuse fraction of H$\alpha$ and continuum UV 
\citep{2000ApJ...541..597H,2003ApJ...586..902H,2005ApJ...619L..67T} and
moreover our selected sources lie in non crowded, low density regions, so leaking might
be more pronounced than close to giant molecular clouds on the arms.

\begin{figure}
\plottwo{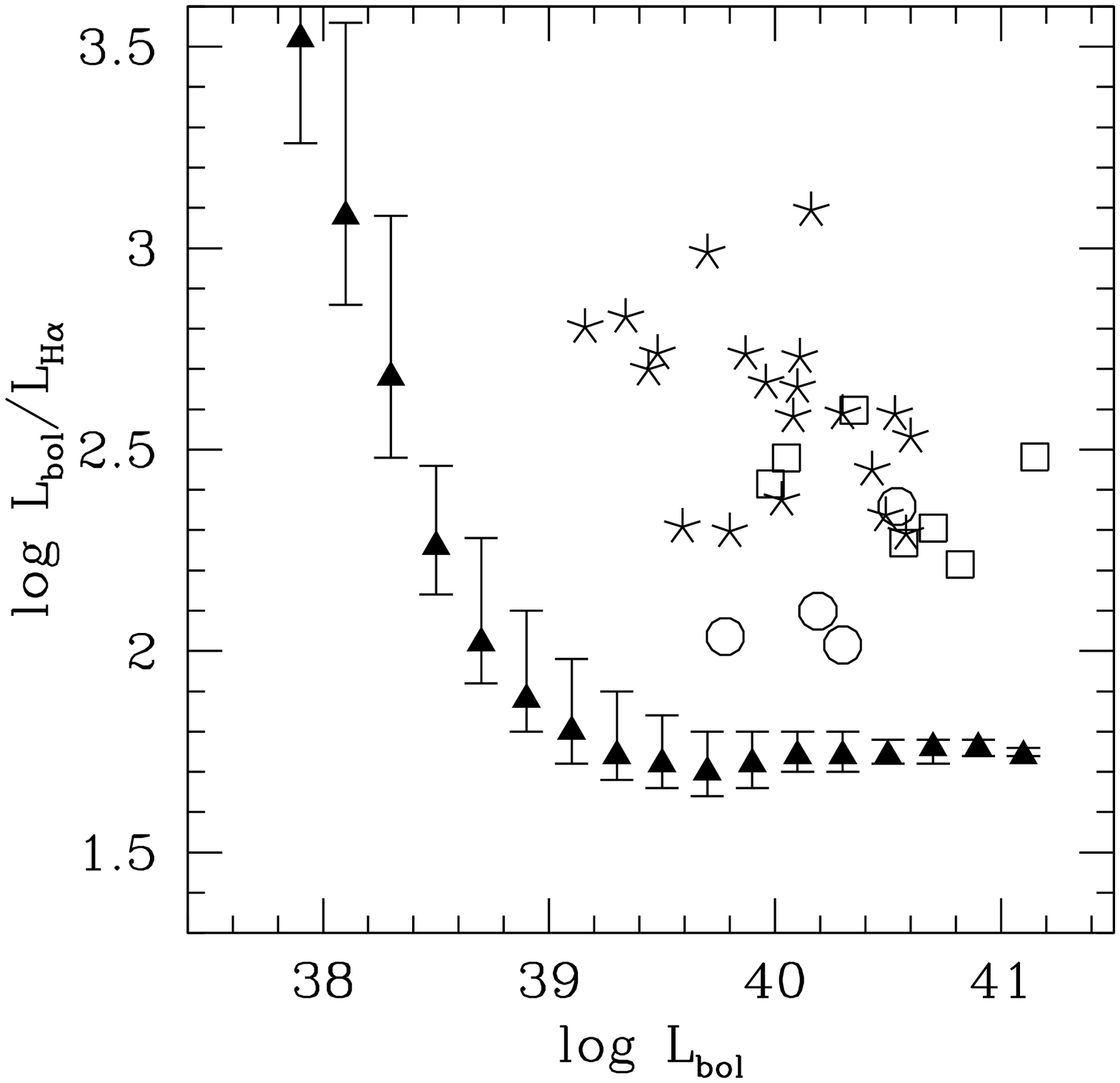}{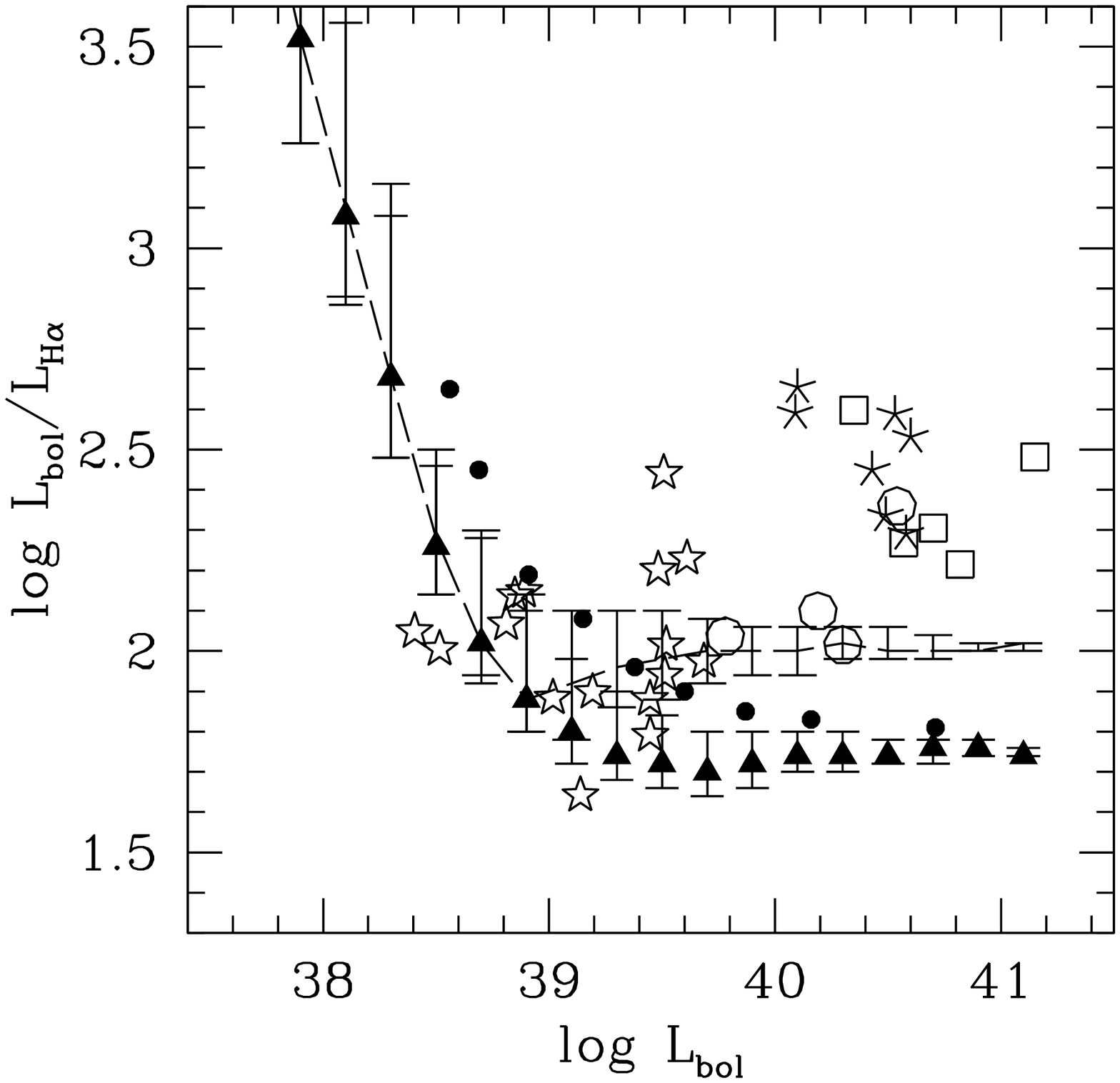}
\caption{The cluster birthline for the stochastically sampled IMF which extends up to
120~M$_\odot$ (filled triangles) and to 40~M$_\odot$ (dashed line in $(b)$).
Data for infrared selected clusters in M33 are shown in $(a)$ when L$_{bol}$ is determined
from SED model fits using a complete IMF up to 120~M$_\odot$. Asterisk symbols are for
clusters younger than 3~$10^6$~yrs, open squares for clusters with age between
3~$10^6$ and 6.3~$10^6$~yrs and open circles for older clusters. In panel $(b)$ we 
compute L$_{bol}$ by adding the emission in the various bands (open star symbols), and
only when this is close to 10$^{40}$~erg~sec$^{-1}$ or above it we show L$_{bol}$
as from SED model fits.  Filled dots trace the birthline for the maximum mass case.
\label{fig:4}
}
\end{figure}
 
Leakage becomes less pronounced if massive stars are missing. In Figure~4$(b)$ we 
show the birthline for a stochastically sampled IMF which extend only up to 40~M$_\odot$ 
as well as the birthline relative to the maximum mass case.
M33 clusters are  compatible with the
two stochastically sampled universal IMF while there are a few clusters that lie well 
below the maximum mass birthline and hence the relative IMF model is less suitable.
However, a universal lower mass cut-off to the IMF for our sample does not solve
the age problem i.e. that younger clusters lie close to the birthline than
older cluster do. This can happen if  
a delayed formation of massive stars is taking place in the low density regions where 
the star formation timescale is longer.

\subsection{The embedded sample}

In order to test a cluster birthline it is important to select newborn clusters over
a wide range of luminosities. 
Cluster ages determined through SED model fits with a fully sampled IMF are reliable
if clusters are luminous enough to be on the flat part of the birthline. For the rising
part of the birthline, one can hope that infrared selected clusters are young if they
are still embedded in molecular gas. Through deep observations of the CO J=1-0
and J=2-1 line we have 14 young clusters candidates which we use to test the
rising part of the birthline. They are dim infrared sources with no or little flux
in the UV and we compute L$_{bol}$ by summing the emission in the various bands.
Figure 5 shows how well these sources trace the birthline for a stochastically
sampled universal IMF.

\begin{figure}[ht!]
\includegraphics[bb=18 104 480 550,scale=0.6]{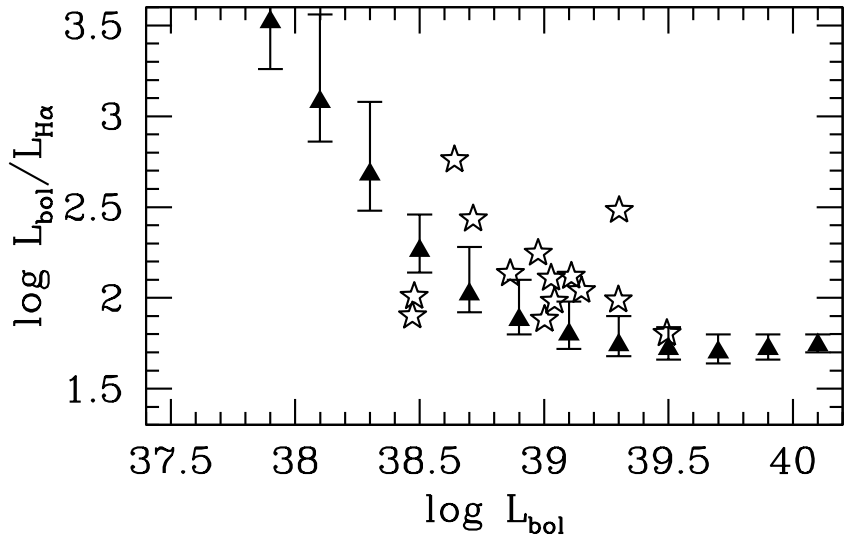}
\caption{The cluster birthline for the stochastically sampled IMF (filled triangles) and
data for infrared selected clusters in M33 embedded or close to a molecular cloud
(open star symbols).
\label{fig:5}
}
\end{figure}

\section{Conclusions}

We have used the cluster birthline, which is the place in the 
log L$_{bol}$ -- log L$_{bol}$/L$_{H\alpha}$ plane where newborn young stellar clusters 
lie, to test the IMF of individual clusters in the Local Group galaxy M33.
Cluster aging or ionizing photon leakage moves the clusters above the birthline.
A stochastically sampled universal IMF with a Salpeter slope at its high mass end 
gives a cluster birthline in agreement with M33 data. Low luminosity clusters, which lie
on the rising part of the birthline, are in better agreement with a stochastic 
universal IMF rather than with truncated IMF models whose maximum stellar mass
depends on cluster mass. 

Fits to the spectral energy distribution of individual
luminous clusters suggest a delayed formation of massive stars i.e. a non
instantaneous burst of star formation in low density environments. This helps
explaining the unexpected result that older clusters lie closer to the birthline
than younger clusters do. The alternative is to invoke a substantial leakage of
ionizing photons in young luminous clusters, in addition to multiple 
burst models.


\bibliography{corbelli_e}

\end{document}